\documentclass[]{JHEP3} 

\usepackage{epsfig,multicol}
\usepackage[square, comma, sort&compress]{natbib}

\usepackage{graphicx}
\usepackage{dcolumn}
\usepackage{bm}
\usepackage{epsfig}

\newcommand\fverb{\setbox\pippobox=\hbox\bgroup\verb}
\newcommand\fverbdo{\egroup\medskip\noindent%
                        \fbox{\unhbox\pippobox}\ }
\newcommand\fverbit{\egroup\item[\fbox{\unhbox\pippobox}]}
\newbox\pippobox

\newcommand{\beqn}{\begin{eqnarray}}
\newcommand{\eeqn}{\end{eqnarray}}
\newcommand{\beqns}{\begin{eqnarray*}}
\newcommand{\eeqns}{\end{eqnarray*}}

\newcommand{\beq}{\begin{equation}}
\newcommand{\eeq}{\end{equation}}
\newcommand{\beqa}{\begin{eqnarray}}
\newcommand{\eeqa}{\end{eqnarray}}

\newcommand{\nn}{\nonumber \\}
\newcommand{\ve}{\varepsilon}
\newcommand{\e}{\epsilon}
\newcommand{\be}{\begin{equation}}
\newcommand{\ee}{\end{equation}}
\newcommand{\ba}{\begin{eqnarray}}
\newcommand{\ea}{\end{eqnarray}}

\def\za{\cal{N}}
\def\za#1.#2{\left\langle#1 \hskip .15 mm #2\right\rangle}

\newcommand{\eps}{\epsilon}

\def\cg{c_\Gamma}

\title{On the Numerical Evaluation of One-Loop Amplitudes: the Gluonic Case}
\author{W.~T.~Giele\footnote{email: giele@fnal.gov}\\
  Fermilab, Batavia, IL 60510, USA}

\author{G.~Zanderighi\footnote{email: g.zanderighi1@physics.ox.ac.uk}\\
   The Rudolf Peierls Centre for Theoretical Physics, 1 Keble Road, University of
   Oxford, UK.}

\preprint{
OUTP-08-07P\\
FERMILAB-PUB-08-119-T}

\abstract{We develop an algorithm of polynomial complexity for
evaluating one-loop amplitudes with an arbitrary number of external
particles.  The algorithm is implemented in the {\bf Rocket}
program. Starting from particle vertices given by Feynman rules, tree
amplitudes are constructed using recursive relations. The tree
amplitudes are then used to build one-loop amplitudes using an integer
dimension on-shell cut method.  As a first application we considered
only three and four gluon vertices calculating the pure gluonic
one-loop amplitudes for arbitrary external helicity or polarization
states.  We compare our numerical results to analytical results in the
literature, analyze the time behavior of the algorithm and the
accuracy of the results, and give explicit results for fixed phase
space points for up to twenty external gluons.}


\begin{document}

\section{Introduction}
The current Tevatron collider and the upcoming Large Hadron Collider
(LHC) experiments require a good understanding of the Standard Model
signals to carry out a successful search for the Higgs particle and
physics beyond the Standard Model. At these hadron colliders QCD plays
an essential role and modeling all of the aspects of the events is
crucial. From the lessons learned at the Tevatron, we need fixed order
calculations matched with parton shower Monte Carlo's and
hadronization models for a successful understanding of the observed
events.

Fixed order calculations are the first step in modeling the events. In
this paper we describe an algorithm for the automated calculation of
one-loop amplitudes.  A successful implementation of numerical
algorithms for evaluating fixed order amplitudes needs to take into
account the so-called complexity of the algorithm. That is, how does
the evaluation time grow with the number of external particles. An
algorithm of polynomial complexity is highly
desirable~\cite{Bern:2008ef}.  Another consideration is the
suitability of the method within a numerical context.  In particular
algebraic methods can be successfully implemented in efficient and
robust algorithms. This can lead to rather different methods from what
one would develop and use in analytic calculations.

Leading order parton level generators are well understood.  These have
been constructed using algebraic manipulation programs to calculate
the tree amplitudes directly from Feynman
diagrams~\cite{MADGRAPH,CompHEP,AMEGIC,Belanger:2003sd,Boos:2004kh}.
However, such a direct approach leads to algorithms of double
factorial complexity.  Techniques such as helicity
amplitudes~\cite{hel1,hel2,hel3,hel4}, color ordering~\cite{col1,col2}
and recursion
relations~\cite{Berends:1987me,Caravaglios:1995cd,Draggiotis:1998gr,BCFrecursion,BCFW}
have been developed and successfully used both in analytic and
numerical calculations of leading order amplitudes. These techniques
all aim at further factorizing the calculation in smaller subsets.  Of
these the recursion relation technique stands out as it maximizes the
factorizability of tree amplitudes and consequently can be implemented
as a polynomial complexity algorithm of rank
four~\cite{Kleiss:1988ne}.  Furthermore, recursion relations have a
simple algebraic structure ideal for numerical evaluation of the tree
amplitudes.  Such a recursive algorithm was successfully implemented
in tree amplitude generators~\cite{ALPGEN,HELAC}.

Compared to leading order generators the status of next-to-leading
order generators is far less advanced.  Calculations based on explicit
one-loop Feynman diagrams using tensor form factor reductions have
been used successfully for four, five and six point
amplitudes. Examples of recent explicit calculations of six point
one-loop amplitudes using such brute force methods are
six photons~\cite{Binoth:2007ca,Ossola:2007bb},
six gluons~\cite{Ellis:2006ss} and $e^+e^-\rightarrow$ four
fermions~\cite{denner1,denner2}. These direct approaches suffer from
worse than factorial complexity.

Alternatively, powerful analytic methods have been developed for
one-loop amplitudes based on generalized unitarity
methods~\cite{Bern:1994zx,Bern:1995db,Bern:1996je,Bern:1996ja,Britto:2004nc,Britto:2004nj,Bern:2005hs,
Bern:2005ji,Bern:2005cq,Berger:2006ci,Britto:2005ha,Britto:2006sj,Mastrolia:2006ki,Anastasiou:2006gt,
Anastasiou:2006jv,Britto:2006fc,Britto:2007tt,Britto:2008vq,Britto:2008sw,Forde:2007mi}.
These methods have been used in analytic calculations leading to
compact expressions for processes such as for example vector boson
plus four partons~\cite{Bern:1996ka,Zqqgg}.  The analytic methods
developed are again based on the principle of factorization. By using
generalized unitarity one can factorize the one-loop amplitude into
tree amplitudes from which the coefficients of the master integrals
can be constructed.

The main problems of converting the on-shell analytic method into
numerical algorithms are twofold. The first issue is resolving
overlapping contributions between quadruple, triple and double
cuts. Using the methods developed in~\cite{OPP,EGK} one can
disentangle the contributions in an algebraic manner.  Secondly, the
numerical implementation of the dimensional regularization within an
on-shell method leads to complications. The on-shell cut lines give
intermediate on-shell particles in non-integer number of dimensions.
These lines are ill-defined and require special considerations. In
particular from the aspect of a numerical implementation one needs a
well-defined method for calculating the contributions originating from
dimensional regularization. Several methods to calculate these
contributions applicable for numerical implementation have been
developed~\cite{Bern:2005cq,Berger:2006ci,CutTools,GKM,OPP2}.  These
developments now allow the construction of numerical algorithms for
evaluating one-loop amplitudes with a large number of external
particles~\cite{Binoth:2008kt,Berger:2008sj}.

In this paper we present an algorithm of rank nine polynomial
complexity to calculate one-loop amplitudes with an arbitrary number
of particles based on the purely algebraic methods developed in
refs.~\cite{EGK,GKM}.  As a first step we implement the algorithm in
the program {\bf Rocket}\footnote{From the Italian {\bf Rucola}: {\bf
R}ecursive {\bf U}nitarity {\bf C}alculation of {\bf O}ne-{\bf L}oop
{\bf A}mplitudes.}  using three and four gluon vertices to calculate
all the helicity amplitudes of the pure gluonic scattering amplitudes.

The outline of the paper is as follows. In section~\ref{sec:method} we
review the method and outline the construction of the required
orthonormal basis vectors and $D$-dimensional polarization vectors.
Section~\ref{sec:results} contains our results: we study the accuracy
of the method, the power-like growth of the computation time and give
explicit single event results for various helicity configurations for
up to twenty gluons. Finally, in section~\ref{sec:conclu} we draw our
conclusions and give an outlook on our future plans.

\section{Construction of the Algorithm}
\label{sec:method}
We implement the methods developed in refs.~\cite{EGK,GKM} with some
minor modifications into the {\bf Rocket} program.
These methods build upon the formalism of ref.~\cite{OPP} by removing
the requirement of the four dimensional spinor language, thereby
allowing for the extension of the method to $D$-dimensional cuts.

To calculate the full one-loop $N$-gluon amplitude, it is sufficient
to be able to calculate the leading color ordered one-loop amplitude.
From these color ordered amplitudes the full one-loop amplitude can be
constructed~\cite{Bern:1990ux,Bern:1994zx}. Eventually the one-loop
amplitude has to be contracted with the tree amplitude, summed over
the colors and integrated over phase space.  The number of different
orderings one needs to calculate can be drastically reduced by noting
that the phase space integration will symmetrize the final state
gluons.  Alternatively, a more physical approach can be employed by
not only randomly sampling over the helicities but also over the
colors of the external gluons.  In the following we will therefore
focus on the leading color ordered amplitudes
$A_N^{[1]}(1,2,\ldots,N)$.

We will use the master integral basis decomposition derived in
ref.~\cite{GKM}. This decomposition in an overcomplete set of master
integrals makes the loop momentum dependence on the dimensionality explicit:
\beqa\label{MasterDecomp}
&& {A}_N^{[1]}=
-\sum_{[i_1|i_5]}\frac{(D-4)}{2}\, e^{(2,0)}_{i_1i_2i_3i_4i_5}\ 
I^{(D+2)}_{i_1i_2i_3i_4i_5} \nn
&&+\sum_{[i_1|i_4]} \left(
d^{(0,0)}_{i_1i_2i_3i_4}\ I^{(D)}_{i_1i_2i_3i_4}
-\frac{(D-4)}{2}\, d^{(2,0)}_{i_1i_2i_3i_4}\ I^{(D+2)}_{i_1i_2i_3i_4}
+\frac{(D-4)(D-2)}{4}\, d^{(4,0)}_{i_1i_2i_3i_4}\ I^{(D+4)}_{i_1i_2i_3i_4}
\right)\nonumber \\
&&+\sum_{[i_1|i_3]} \left(c^{(0,0)}_{i_1i_2i_3}\ 
I^{(D)}_{i_1i_2i_3}-\frac{(D-4)}{2}c^{(2,0)}_{i_1i_2i_3}\ 
I^{(D+2)}_{i_1i_2i_3} \right)\nn
&&+\sum_{[i_1|i_2]} \left( b^{(0,0)}_{i_1i_2}\ 
I^{(D)}_{i_1i_2}-\frac{(D-4)}{2}b^{(2,0)}_{i_1i_2}\ 
I^{(D+2)}_{i_1i_2}\right)\ ,
\eeqa
where we introduced the short-hand notation $[i_1|i_n] = 1\leq
i_1<i_2<\cdots<i_n \leq N$ and
\beq
I^D_{i_1,\ldots i_N} = \int \frac{d^Dl}{i \pi^{D/2}}
\frac{1}{d_{i_1}d_{i_2}\ldots d_{i_N}}\,,
\eeq
with $d_{i}= d_i(l)=(l+q_i)^2=(l+p_1+\cdots+p_i)^2$. 
In this basis the coefficients
$b^{(0,0)}_{i_1i_2}$, $b^{(2,0)}_{i_1i_2}$, $c^{(0,0)}_{i_1i_2i_3}$,
$c^{(2,0)}_{i_1i_2i_3}$, $d^{(0,0)}_{i_1i_2i_3i_4}$,
$d^{(2,0)}_{i_1i_2i_3i_4}$, $d^{(4,0)}_{i_1i_2i_3i_4}$, and
$e^{(2,0)}_{i_1i_2i_3i_4i_5}$ are independent of the loop momentum
dimensionality $D$.  Because the coefficients
$d^{(4,0)}_{i_1i_2i_3i_4}$, $c^{(2,0)}_{i_1i_2i_3}$ and
$b^{(2,0)}_{i_1i_2}$ are multiplied with a dimensional factor $(D-4)$
they cannot be determined using four dimensional cuts\footnote{Note
that the terms proportional to $e^{(2,0)}_{i_1i_2i_3i_4i_5}$ and
$d^{(2,0)}_{i_1i_2i_3i_4}$ will contribute only at ${\cal
O}(\epsilon)$.}.  Therefore we need to extend the dimensionality of the
cut line to higher dimensions. We choose the dimensionality to be an
integer, resulting in a well defined on-shell particle after performing the
cut~\cite{GKM}.

By applying quintuple, quadruple, triple and double $D_s$-dimensional
cuts (where $D_s\geq D$ denotes the dimensionality of the spin-space)
we can determine the coefficients of the parametric form of the
one-loop amplitude.  This requires the calculation of the factorized
un-integrated one-loop amplitude
\beqa\label{factorize}
\mbox{Res}_{i_1\cdots i_M}({\cal A}^{[1]}_N(l))
&=&\Big(d_{i_1}\times\cdots\times d_{i_M}\times {\cal
A}_N^{[1]}(l)\Big)_{d_{i_1}=\cdots =d_{i_M}=0}
\nonumber\\
&=&\sum_{\{\lambda_1,\ldots,\lambda_M\}=1}^{D_s-2}
\left(\prod_{k=1}^M {\cal A}_{i_{k+1}-i_k}^{[0]}(l_{i_k}^{(\lambda_k)},p_{i_k+1},\ldots,p_{i_{k+1}},-l_{i_{k+1}}^{(\lambda_{k+1})})\right),
\eeqa
where $M \leq 5$ and the $D$-dimensional loop momentum $l$ has to be
chosen such that $d_{i_1}(l)=\cdots =d_{i_M}(l)=0$.  As a result of
the on-shell condition, the tree amplitudes ${\cal A}^{[0]}$ have in
addition to the external four dimensional gluons, two
$D_s$-dimensional gluons with complex momenta.  Dimensional
regularization requires that $D_s\geq D$ such that the ultra-violet
poles are regulated. These higher dimensional gluons have $(D_s-2)$
polarization states.  To calculate these tree amplitudes we use the
standard Berends-Giele recursion relation~\cite{Berends:1987me} which
is valid in arbitrary dimension and for complex momenta.

The generic solution for the loop momentum in eq.~(\ref{factorize}) is
given by
\beq\label{lsolution}
l^{\mu}_{i_1\cdots i_M}=V^{\mu}_{i_1\cdots i_M}
+\sqrt{\frac{-V_{i_1\cdots i_M}^2
}{\alpha_M^2+\cdots+\alpha_D^2}}\left(\sum_{i=M}^{D}
\alpha_i\,n_i^{\mu}\right)\,,
\eeq
for arbitrary values of the variables $\alpha_i$.\footnote{The
conformal transformation applied on the coefficients $\alpha_i$ is
only valid when $V_{i_1\cdots i_M}^2\neq 0$. For the special case
$V_{i_1\cdots i_M}^2= 0$ a similar solution is found.}

The vector $V^{\mu}_{i_1\cdots i_m}$ is defined in the space spanned
by the denominator offset momenta $\{q_{i_1},\ldots,q_{i_M}\}$, while
the orthonormal basis vectors $\{n_M^{\mu},\ldots,n_D^{\mu}\}$ span
the space orthogonal to the space spanned by these
momenta~\cite{EGK,GKM}.
Given the solution to the on-shell conditions $l^{\mu}_{i_1\cdots
i_M}$ in eq.~(\ref{lsolution}), the loop momenta flowing into the tree
amplitudes $l_{i_k}$ and $l_{i_{k+1}}$ in eq.~(\ref{factorize}) are
fixed by momentum conservation (see ref.~\cite{EGK}).

Finally we need a stable and general method for constructing the
orthonormal set of $(D-M)$ basis vectors and the $(D_s-2)$
polarization vectors.  To accomplish this we use the generalized
Kronecker delta tensors~\cite{van Oldenborgh:1989wn} given by
\beq
\delta^{\mu_1\mu_2\cdots\mu_R}_{\nu_1\nu_2\cdots\nu_R}=\left| \begin{array}{cccc} 
\delta_{\nu_1}^{\mu_1} & \delta_{\nu_2}^{\mu_1} & \dots & \delta_{\nu_R}^{\mu_1} \\
\delta_{\nu_1}^{\mu_2} & \delta_{\nu_2}^{\mu_2} & \dots & \delta_{\nu_R}^{\mu_2} \\
\vdots & \vdots & &\vdots \\ 
\delta_{\nu_1}^{\mu_R} & \delta_{\nu_2}^{\mu_R} & \dots & \delta_{\nu_R}^{\mu_R}
\end{array}\right|\ .
\eeq
We use the notation
\beq
\delta^{p\mu_2\cdots\mu_R}_{\nu_1q\cdots\nu_R}\equiv
\delta^{\mu_1\mu_2\cdots\mu_R}_{\nu_1\nu_2\cdots\nu_R}p_{\mu_1}q^{\nu_2}\ .
\eeq
The $(R-1)$-particle Gram determinant is then given by
\beq
\Delta(k_1,k_2,\cdots,k_{R-1})=\delta^{k_1k_2\cdots k_{R-1}}_{k_1k_2\cdots k_{R-1}}\ .
\eeq
Note that for $R\geq D+1$ the generalized Kronecker delta is zero. For
the special case $R=D$ we have the factorization of the Kronecker
delta into a product of Levi-Civita tensors:
$\delta^{\mu_1\mu_2\cdots\mu_
D}_{\nu_1\nu_2\cdots\nu_D}=\varepsilon^{\mu_1\mu_2\cdots\mu_D}\varepsilon_{\nu_1\nu_2\cdots\nu_D}$.

Given a set of momenta $\{q_1,q_2,\ldots,q_M\}$ in a $D$-dimensional
space-time we describe here how to construct the orthonormal set of
basis vectors $\{n_{1},\ldots,n_{D-M}\}$ such that $q_i\cdot n_j=0$
and $n_i\cdot n_j=\delta_{ij}$. The set of momenta span the
$M$-dimensional sub-space. The basis vector set $\{n_i\}$ spans the
orthogonal $(D-M)$-dimensional space.
For the $k$-th basis vector we choose the arbitrary vector $b_k$. The
vector is then given by
\beq
\label{BasisVectors}
n_k^{\mu}=\frac{\delta_{b_k b_{k-1}\cdots b_1q_1\cdots
q_M}^{\mu\phantom{_k} b_{k-1}\cdots b_1q_1\cdots q_M}}
{\sqrt{\Delta(b_{k-1},\ldots,b_1,q_1,\ldots,q_M)\Delta(b_k,\ldots,b_1,q_1,\ldots,q_M)}}\
,
\eeq
with $q_i\cdot n_K=0$ and $n_j\cdot n_k=\delta_{ik}$.
Note that arbitrary vector $b_{D-M}$ in the basis vector
$n_{D-M}^{\mu}$ drops out trivially because
$\delta_{\mu_1\cdots\mu_D}^{\nu_1\cdots\nu_D}=\ve_{\mu_1\cdots\mu_D}\times
\ve^{\nu_1\cdots\nu_D}$.  This means that in the construction of the
basis we used $(D-M-1)$ arbitrary vectors.

We can also use the above construction of the orthonormal basis
vectors to define the $D_s$-dimensional polarization states of the cut
gluon lines.  Given a $D_s$-dimensional gluon with light-cone momentum
$p$ we want to construct a set of the $(D_s-2)$ polarization vectors
$\{e_{\mu}^{(i)}\}$. The polarization vectors have the property
$p\cdot e^{(i)}=0$, $e^{(i)}\cdot e^{(j)}=-\delta_{ij}$ and the spin
sum is
\beq\label{PolSum}
\sum_{i=1}^{D_s-2} e_{\mu}^{(i)}e_{\nu}^{(i)}=
-g_{\mu\nu}^{(D_s)}+\frac{p_{\mu}v_{\nu}+p_{\nu}v_{\mu}}{p\cdot v}
-\frac{p_{\mu}p_{\nu}v^2}{(p\cdot v)^2}\ ,
\eeq
where $v_{\mu}$ is the polarization gauge vector. Note that if
$v_{\mu}$ is a light-cone vector the last term on the right-hand side
is zero.  The polarization vectors are easily constructed. We
construct the $(D_s-2)$ orthonormal basis vectors $n_i^{\mu}$ with
respect to the two four-vectors $\{p,v\}$. We now define the $(D_s-2)$
polarization vectors as $e_{\mu}^{(j)}=i\,{n_j}_{\mu}$. Then $p\cdot
e^{(i)}=0$, $e^{(i)}\cdot e^{(j)}=-n_i\cdot n_j=-\delta_{ij}$, and
eq.~(\ref{PolSum}) is satisfied.

Once all coefficients in eq.~(\ref{MasterDecomp}) have been determined
using an appropriate set of cuts and loop momentum solution vectors,
we can algebraically continue the dimensionality to the non-integer
limit: $D\rightarrow 4-2\e$.  This limit can be performed in different
manners, leading to different schemes.  In the four dimensional
helicity scheme~\cite{Bern:1991aq,Bern:2002zk} the continuation is
defined as $D_s\rightarrow 4$, $D\rightarrow 4-2\e$ while $D_s\geq D$.
In the 't Hooft-Veltman scheme~\cite{'t Hooft:1972fi} the limit is
defined as $D_s\rightarrow 4-2\e$, $D\rightarrow 4-2\e$ while $D_s\geq
D$.

Because we are interested in the next-to-leading order contributions
we can neglect terms of order $\e$ in the continuation of the
dimensionality. We then find for the color ordered one-loop amplitude
\beqa
&& {A}_N^{[1]}= \sum_{[i_1|i_4]}
d^{(0,0)}_{i_1i_2i_3i_4}\ I^{(4-2\e)}_{i_1i_2i_3i_4}
+\sum_{[i_1|i_3]} c^{(0,0)}_{i_1i_2i_3}\ I^{(4-2\e)}_{i_1i_2i_3}
+\sum_{[i_1|i_2]} b^{(0,0)}_{i_1i_2}\ I^{(4-2\e)}_{i_1i_2} \nonumber \\
&&-\sum_{[i_1|i_4]}
\frac{d^{(4,0)}_{i_1i_2i_3i_4}}{6}
+\sum_{[i_1|i_3]} \frac{c^{(2,0)}_{i_1i_2i_3}}{2}
-\sum_{[i_1|i_2]}\frac{(q_{i_1}-q_{i_2})^2}{6} b^{(2,0)}_{i_1i_2}
+{\cal O}(\e)\ .
\eeqa
The terms in the first line give rise to the so-called
cut-constructible part of the amplitude~\cite{Bern:1994cg}. The terms
in the second line can be identified with the rational part. In the
approach used in this paper the division between these two
contributions is not relevant.

For the numerical evaluation of the bubble, triangle and box master
integrals we use the package developed in ref.~\cite{Ellis:2007qk}.
%

\section{Results}
\label{sec:results}

The algorithm is implemented in the {\bf Rocket} {\tt Fortran 95}
program. We first perform a series of checks by comparing to existing
analytic results in the literature and by performing internal
consistency checks.  Next the accuracy of the results up to eleven
external gluons is examined by comparing to the analytically known
one-loop $N$-gluon helicity amplitudes.  We then show that the
evaluation time as a function of the number of external gluons is
consistent with a degree nine polynomial.  Finally, we present
explicit results for a few fixed phase space points for selected
helicity configurations with up to twenty external gluons.

In next-to-leading order calculations, the external gluons of the
one-loop amplitude are identified with the momenta of well-separated
jets.  We therefore impose the following set of cuts on the generated
phase space momenta
\beq
|\eta_i | < 3\,,\qquad 
p_{\perp, i} > 0.01 \sqrt{s}\, , \qquad 
R_{ij} = \sqrt{\phi_{ij}^2+\eta_{ij}^2} > 0.4 \,, 
\eeq
where $\eta_i$ and $p_{\perp, i}$ denote the rapidity and transverse
momentum of particle $i$, $\phi_{ij} = |\phi_i - \phi_j|$ and
$\eta_{ij} = |\eta_i - \eta_j|$ denote the distance in azimuthal angle
and rapidity between particle $i$ and $j$ and $s=(E_1+E_2)^2$ is the
center of mass collision energy squared.

The unrenormalized one-loop $N$-gluon results are given in the four
dimensional helicity scheme (FDH).  The conversion to the 't
Hooft-Veltman (HV) scheme is given by
\begin{equation}
{A}_{\rm v, HV} = {A}^v_{\rm v, FDH} - \frac{\cg}{3} {A}_{\rm tree}\,,
\label{THVtoFDH}
\end{equation}
with 
\begin{equation}
\cg = {(4 \pi)^\eps \over 16 \pi^2 }
{\Gamma(1+\eps)\Gamma^2(1-\eps)\over\Gamma(1-2\eps)}\ .
\label{cgdef}
\end{equation}

We will give explicit results for double poles, single poles and the
constant part after extraction of the factor $\cg$.
The pole structure of the $N$-gluon one-loop amplitude is given
by~\cite{LoopPole1,LoopPole2,LoopPole3}
\begin{equation}
\label{eq:poles}
{A}_{\rm v, poles} = c_\Gamma\left(
-\frac{N}{\epsilon^2} +\frac{1}{\epsilon}\left(\sum_{i=1}^N \ln
\left(\frac{-s_{i,i+1}}{\mu^2}\right) - \frac{11}{3}\right)
\right)A_{\rm tree}\,,
\end{equation}
where as usual the indices are modulo $N$.  For the dimensional scale
$\mu$ we use the center of mass collision energy, $\mu^2=s$.

\subsection{Checks of the Results}
We have performed several checks on our numerical implementation: 
\begin{itemize}
\item we checked that the pole structure of the one-loop amplitudes agree with
      eq.~(\ref{eq:poles}) up to twenty gluons;
\item the number of loop momentum solutions to the on-shell
      conditions in eq.~(\ref{lsolution}) is infinite.
      This means we can solve for the coefficients in the
      parametric form of the integrand using many different
      sets of equations. We verified that the results are independent of
      the chosen loop momentum solutions;
\item we checked that the results are independent of all auxiliary
      vectors introduced in e.g. eq.~(\ref{BasisVectors}) used to construct
      both the orthonormal basis and the polarization vectors;
\item we verified that our results are independent of 
      the choice of dimensionality of the cut lines; 
\item for six gluons we compared all one-loop helicity amplitudes with the numerical
      results of~\cite{Ellis:2006ss};
\item for up to twenty gluons we compared with the analytically known 
      one-loop helicity amplitudes where all gluons have
      positive helicities~\cite{Mahlon:1993si,Bern:2005ji};
\item for up to twenty gluons we compared with the analytically known 
      one-loop helicity amplitudes where all
      but one gluon have positive helicities~\cite{Mahlon:1993si,Bern:2005ji};
\item for up to twenty gluons we compared with the analytically known 
      one-loop helicity amplitudes where all
      but two adjacent gluons have positive helicities~\cite{Bern:1994zx,Bern:1994cg,Forde:2005hh}.
\end{itemize}
All checks performed were successful. 

\subsection{Study of the Accuracy}
To study the numerical accuracy of the on-shell method implemented in
{\bf Rocket} we define
\begin{equation}
\varepsilon_C = \log_{10} \frac{|A_{N}^{\rm v, unit}-A_{N}^{\rm v,
anly}|}{|A_{N}^{\rm v, anly}|}\,,
\end{equation}
where ``unit'' denotes the result obtained with the on-shell method and
``anly'' the analytical result for the constant parts of the one-loop helicity
amplitudes (or in the case of $N=6$ the numerical results
of~\cite{Ellis:2006ss}).
Similarly, where relevant, we denote by $\varepsilon_{\rm DP}$ and
$\varepsilon_{\rm SP}$ the accuracy on the double and single poles,
respectively.

\begin{figure}[tbcp]
\begin{center}
\includegraphics[angle=270,scale=0.43]{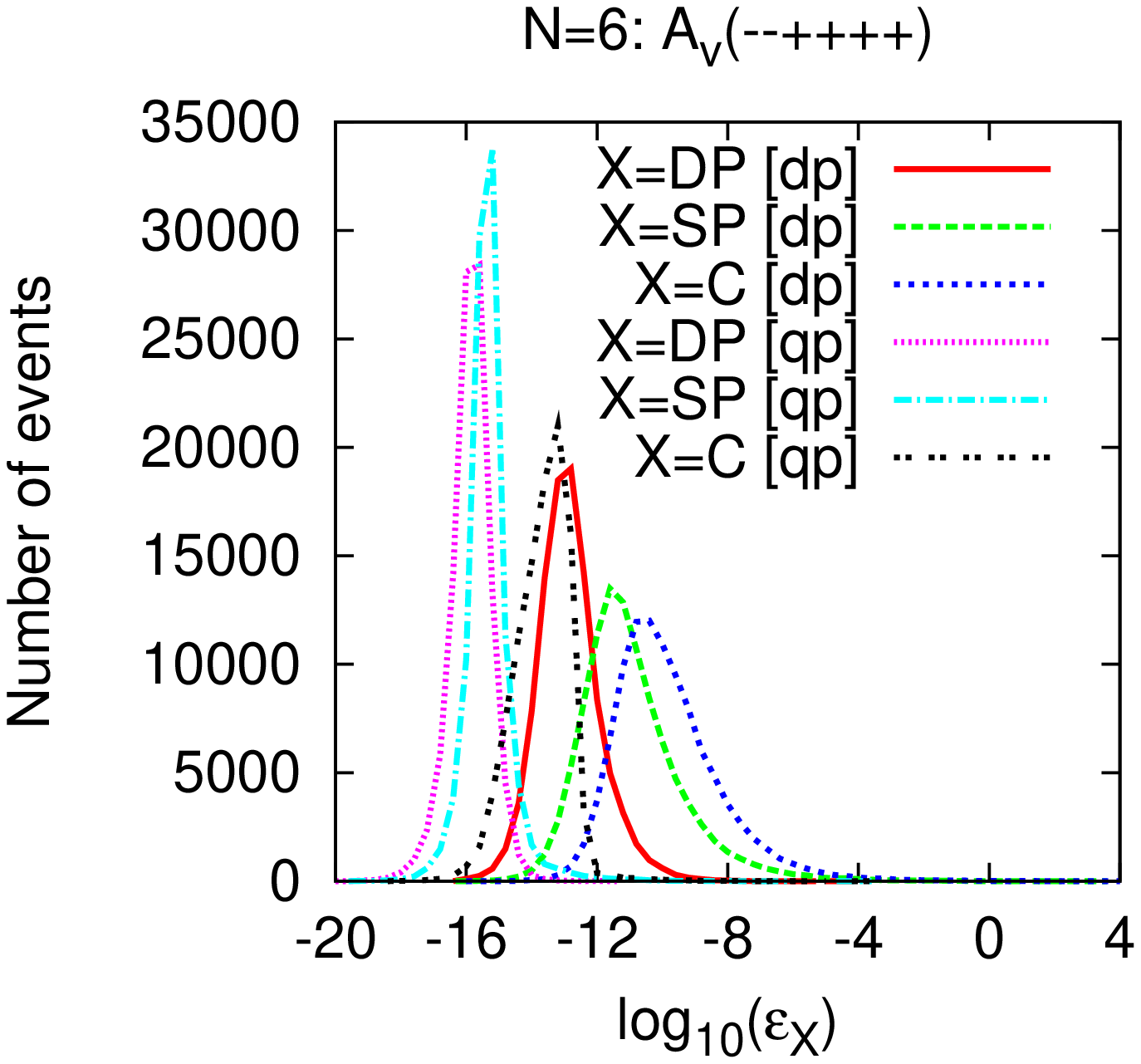}
\includegraphics[angle=270,scale=0.43]{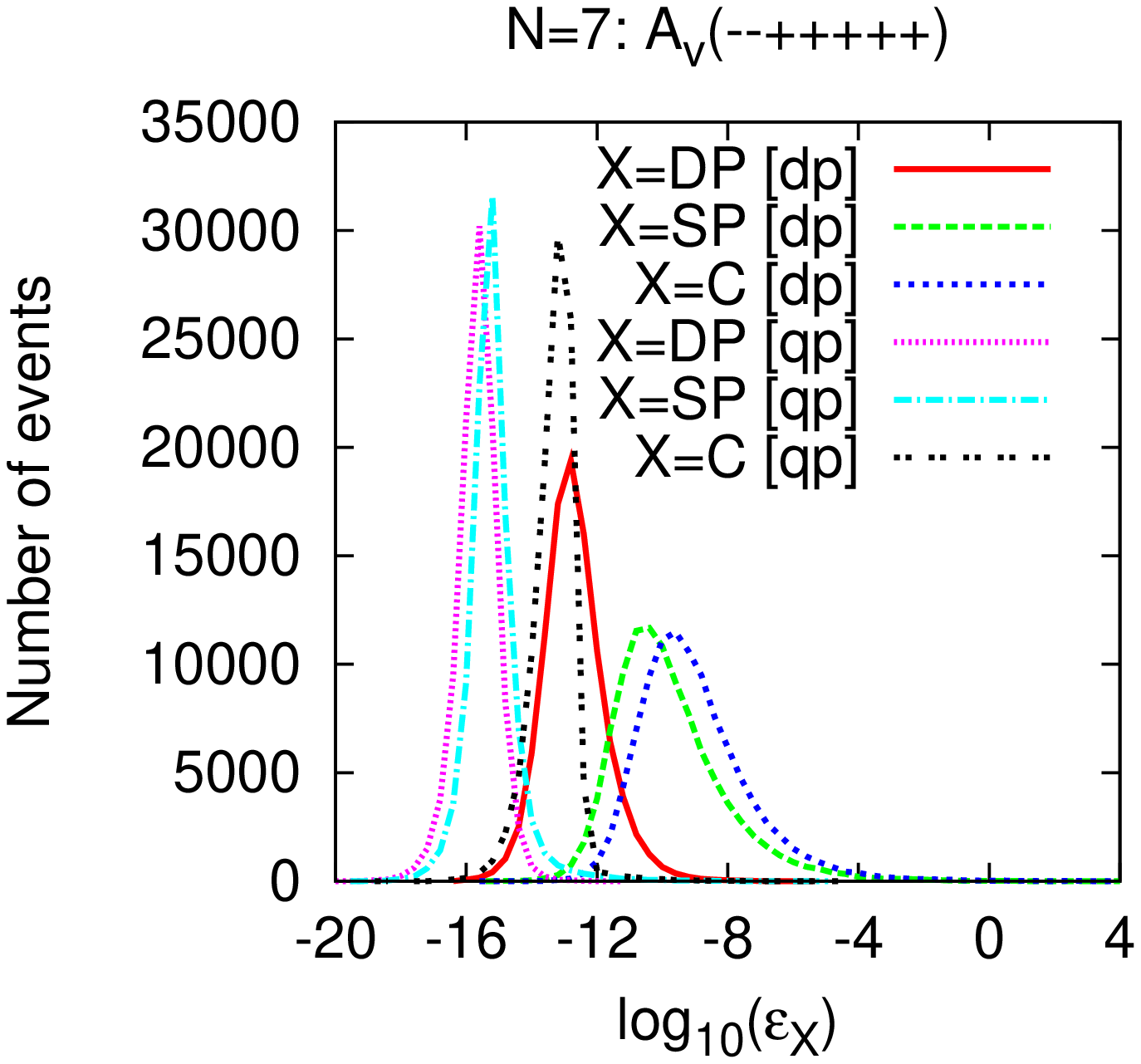}
\includegraphics[angle=270,scale=0.43]{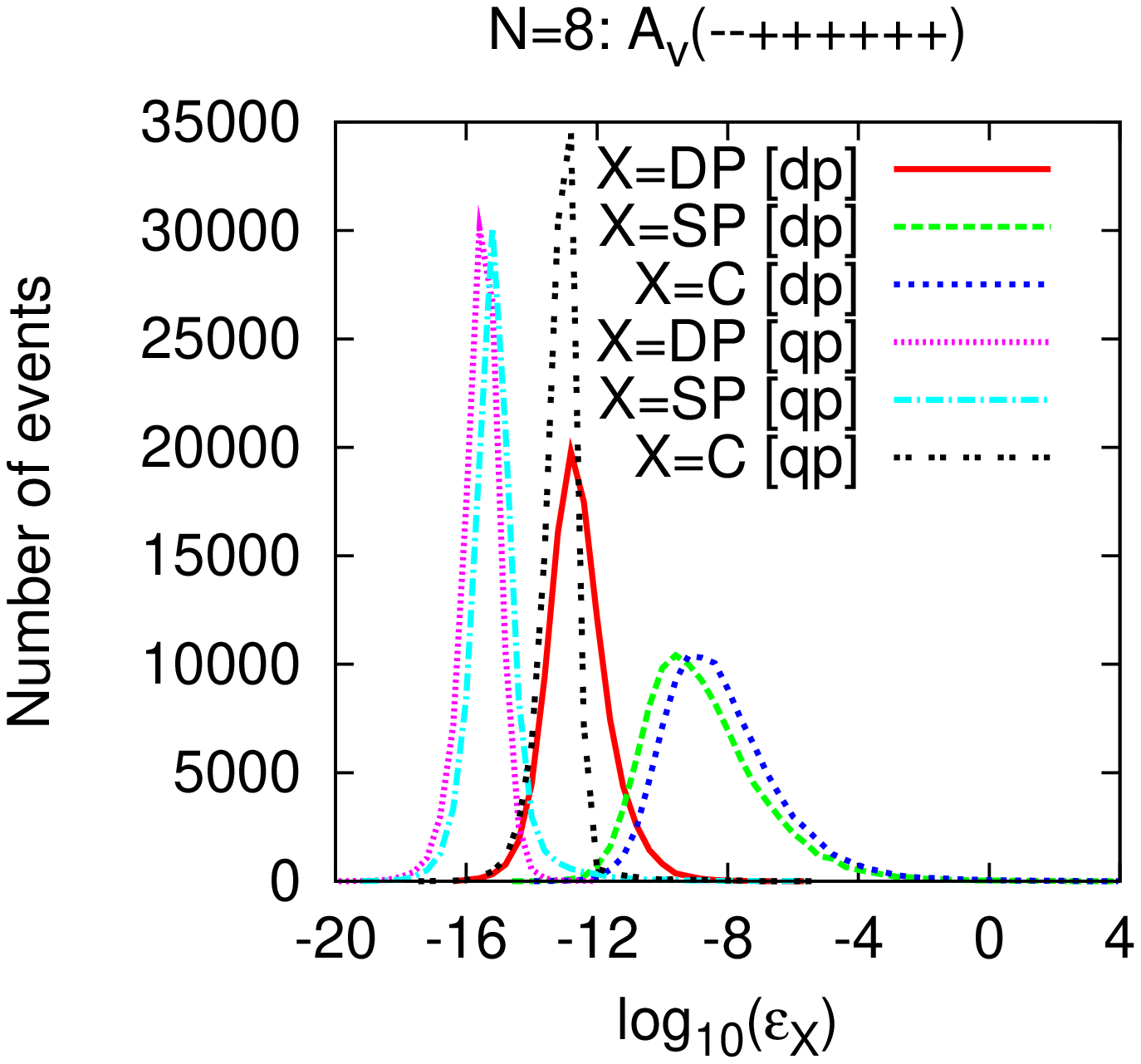}
\includegraphics[angle=270,scale=0.43]{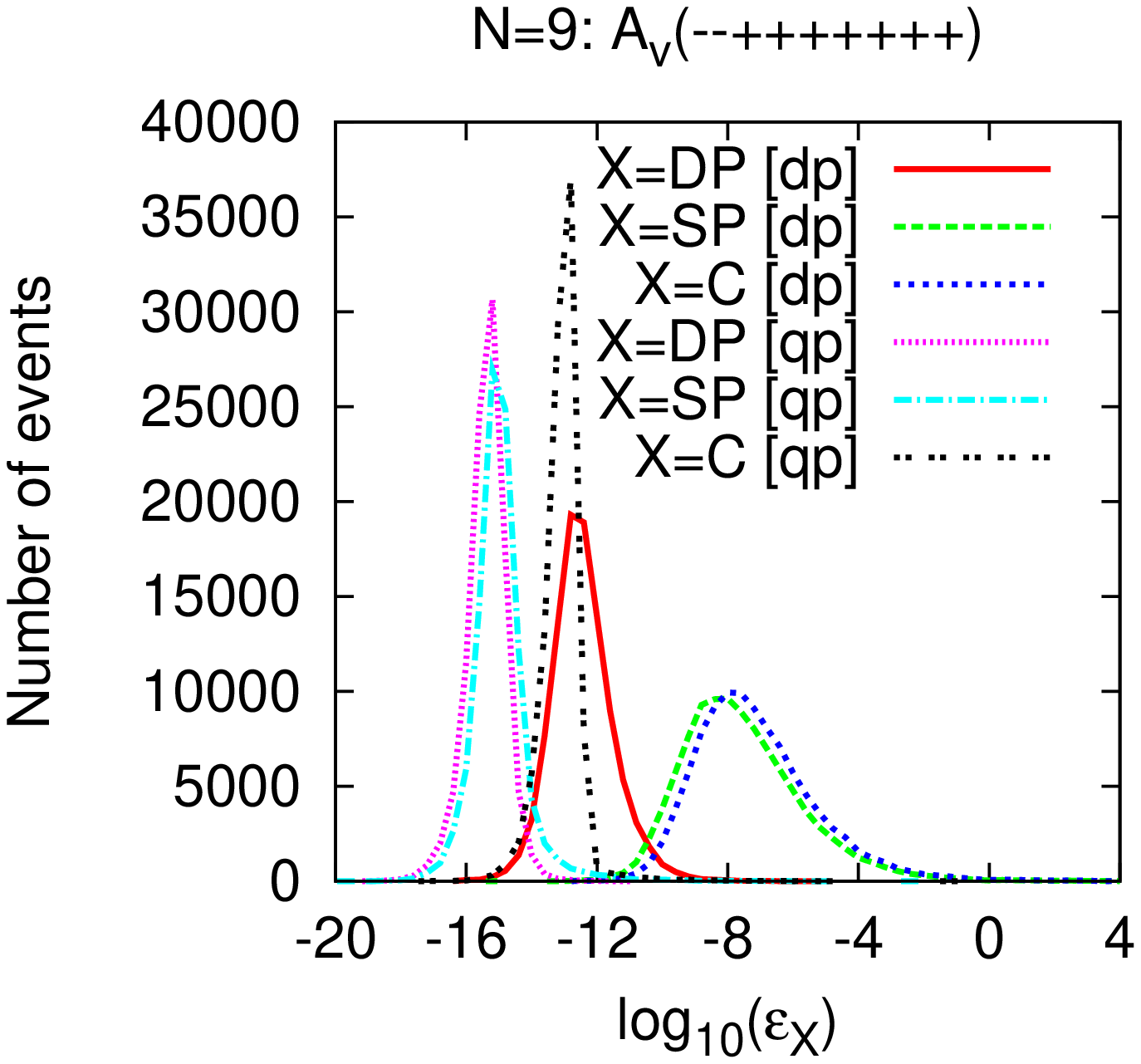}
\includegraphics[angle=270,scale=0.43]{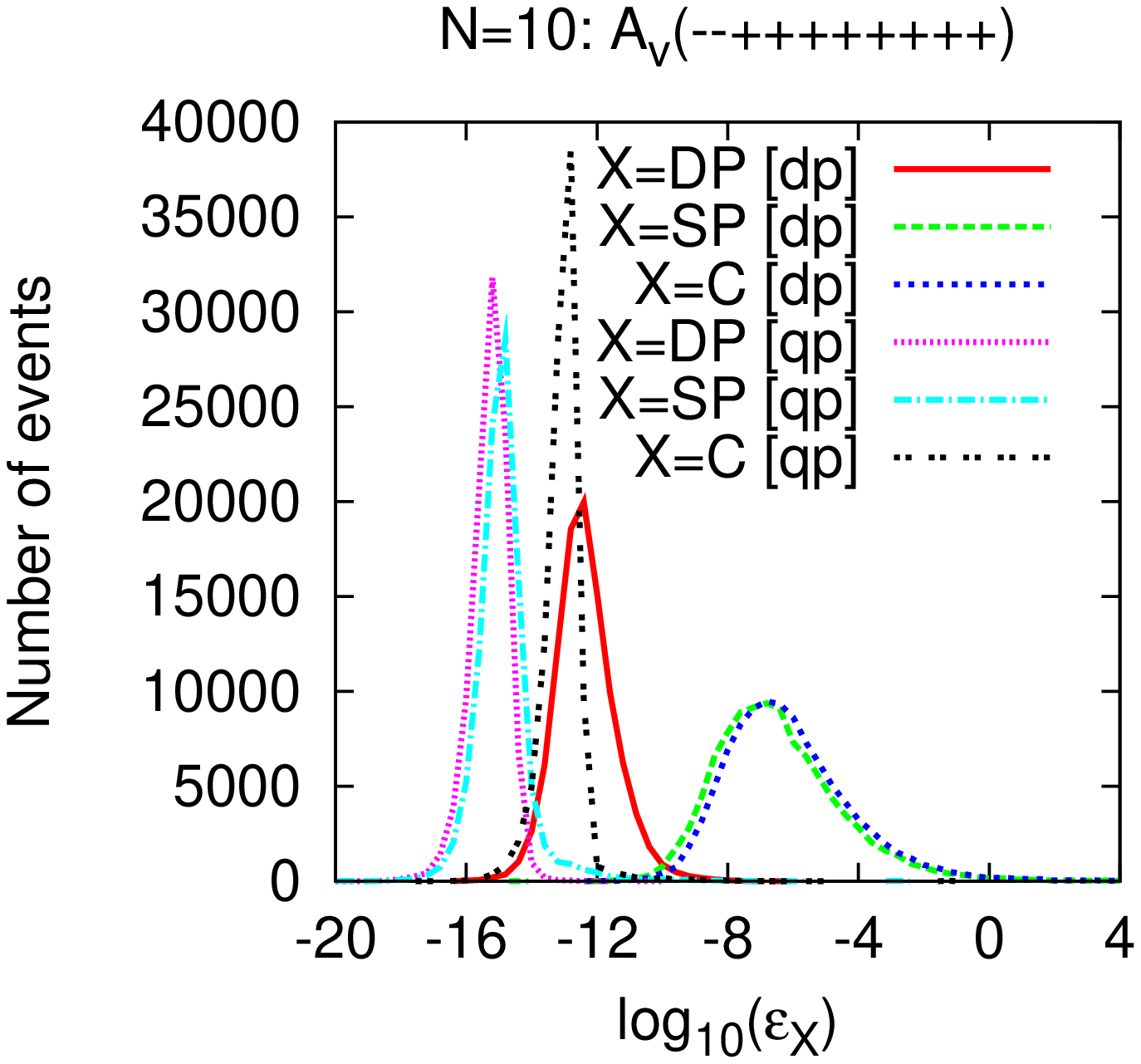}
\includegraphics[angle=270,scale=0.43]{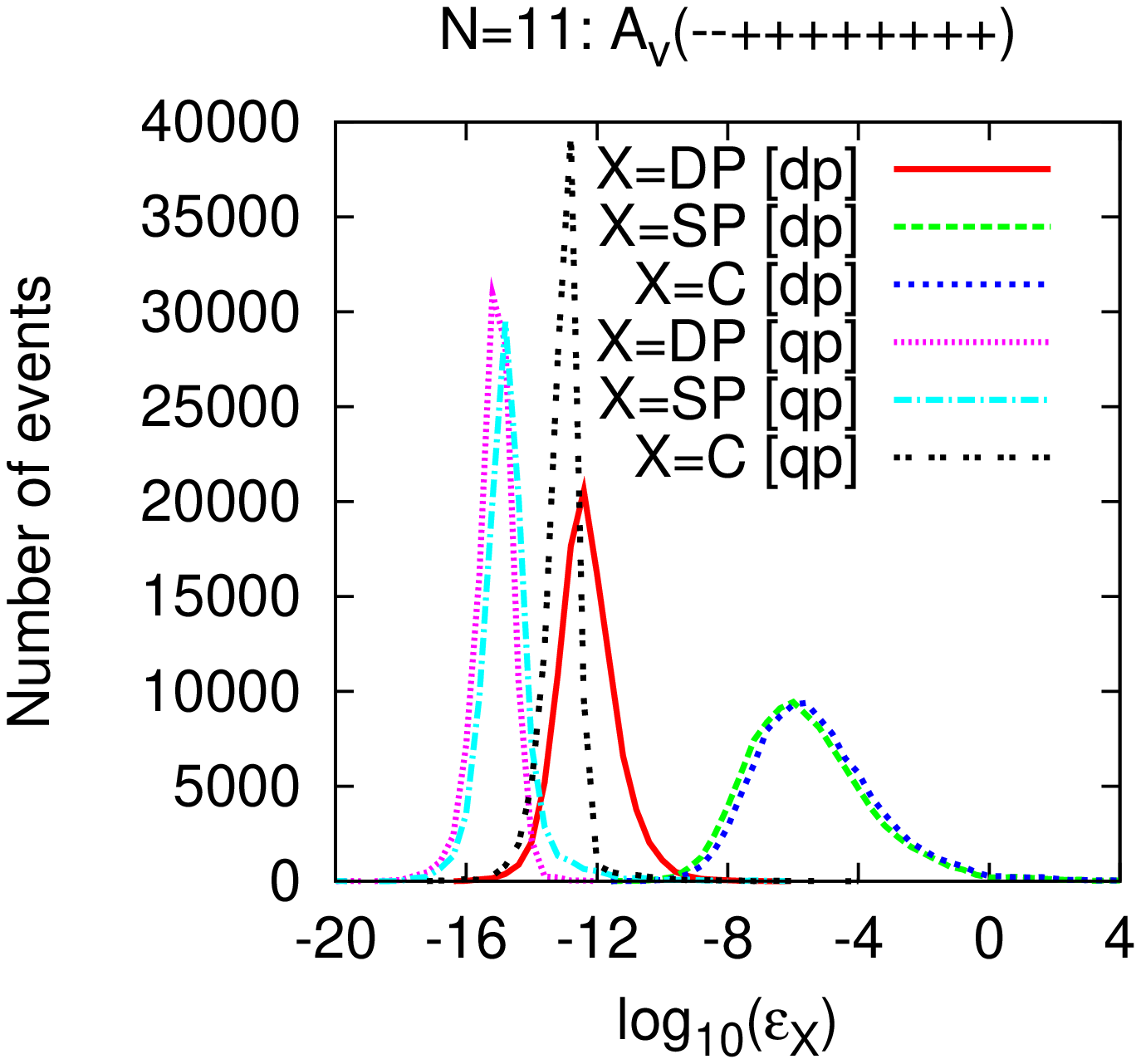}
\end{center}
\caption{Accuracy on the double pole, single pole and constant part of
the MHV amplitude with adjacent negative helicities for 6 up to 11
external gluons. Double ([dp]) and quadrupole ([qp]) precision results for 100,000
phase space points are shown. See text for more details.}
\label{fig:accuracy611}
\end{figure}

In fig.~(\ref{fig:accuracy611}) we show the accuracy for the two
adjacent minus helicity gluon MHV one-loop amplitudes,
$A_N^{[1]}(--+\cdots+)$, for $N$ ranging between six and eleven. The
100,000 phase space points used for each multiplicity are generated
uniformly in phase space using the Rambo
algorithm~\cite{Kleiss:1985gy}.  We plot the accuracy for the double
pole ($\rm X=DP [dp]$, solid, red), the single pole ($\rm X=SP [dp]$,
green, dot-dashed) and the constant part ($\rm X=C [dp]$, blue,
dotted). We first examine the six-gluon plot
(fig.~(\ref{fig:accuracy611}), top left) and see that an excellent
accuracy can be reached for all three contributions, the position of
the peak being at $\epsilon_{\rm DP} = 10^{-12.8}$, $\epsilon_{\rm SP}
= 10^{-11.6}$, and $\epsilon_{\rm C} = 10^{-10.8}$, respectively.
The tail of the distribution reaching to large values of $\epsilon$
contains only a very few points for the single pole and the constant
term.  This lack of agreement is due to numerical instabilities. The
well-known sources of instabilities are vanishing Gram determinants or
other small intermediate denominators.
Several techniques have been developed in the past do deal with such
exceptional points, such as developing systematic
expansions~\cite{Giele:2004ub,Ellis:2005zh,Denner:2005nn} or
interpolating across the singular regions~\cite{DelDuca:2001fn}.
Similarly to what is done
in~\cite{Belanger:2003sd,CutTools,Berger:2008sj}, we adopt here a more
brute force approach and recur to quadrupole precision.  In
(fig.~(\ref{fig:accuracy611}), top left) we see three more curves:
they correspond to the numerical accuracy on the same 100,000 phase
space points when the one-loop amplitude is computed in quadrupole
precision.\footnote{Only the coefficients of the master integrals are
computed in quadrupole precision, master integrals are still
calculated in double precision.}
One sees that the positions of the peaks move even more to the left,
the peak of the double pole is now at $\epsilon_{\rm DP} = 10^{-15.6}$
(magenta, dot-dashed line, labelled $\rm X=DP [qp]$), of the single
poles is at $\epsilon_{\rm SP} = 10^{-15.2}$ (light blue, dot-dashed,
labeled $\rm X=SP [qp]$) and of the constant part is that
$\epsilon_{\rm C} = 10^{-13.2}$ (black, dot-dashed, labelled $\rm X=C
[qp]$). More importantly, out of 100,000 phase space points samples,
not a single point has an accuracy worse than $10^{-4}$.

We can now examine what happens when the number of external gluons is
increased. At double precision we can see from the position of the
peaks and the width of the distributions that the accuracy slowly
worsens with increasing $N$. This is due to a slow accumulation of
errors when more terms are added together and to the fact that there
are potentially more instabilities.
However, at quadrupole precision we see no appreciable worsening of
the accuracy with increasing $N$.
For $N=11$ the peak of the double pole is now at $\epsilon_{\rm DP} =
10^{-15.2}$, of the single poles is at $\epsilon_{\rm SP} =
10^{-14.8}$ and of the constant part is that $\epsilon_{\rm C} =
10^{-12.8}$. Again, out of 100,000 phase space points sampled, not a
single point has an accuracy worse than $10^{-4}$.
Up to $N=11$ (and probably even for
more gluons) quadrupole precision is sufficient to guarantee an
accuracy needed for any next-to-leading order QCD correction.
If higher precision is desired one can choose to evaluate the 
few phase space points which
have insufficient precision using an
arbitrary precision packages such as~\cite{arprec}, at the cost
of higher computation time. 

We note that while the plots here presented are for the MHV amplitudes,
we performed a similar study for the finite amplitudes
($A_N^{[1]}(+\cdots+)$, $A_N^{[1]}(-+\cdots+)$) and
obtain very similar results.
This indicates that the accuracy is
essentially independent of the helicities of the external gluons.

\begin{figure}[t]
\begin{center}
\includegraphics[angle=270,scale=0.49]{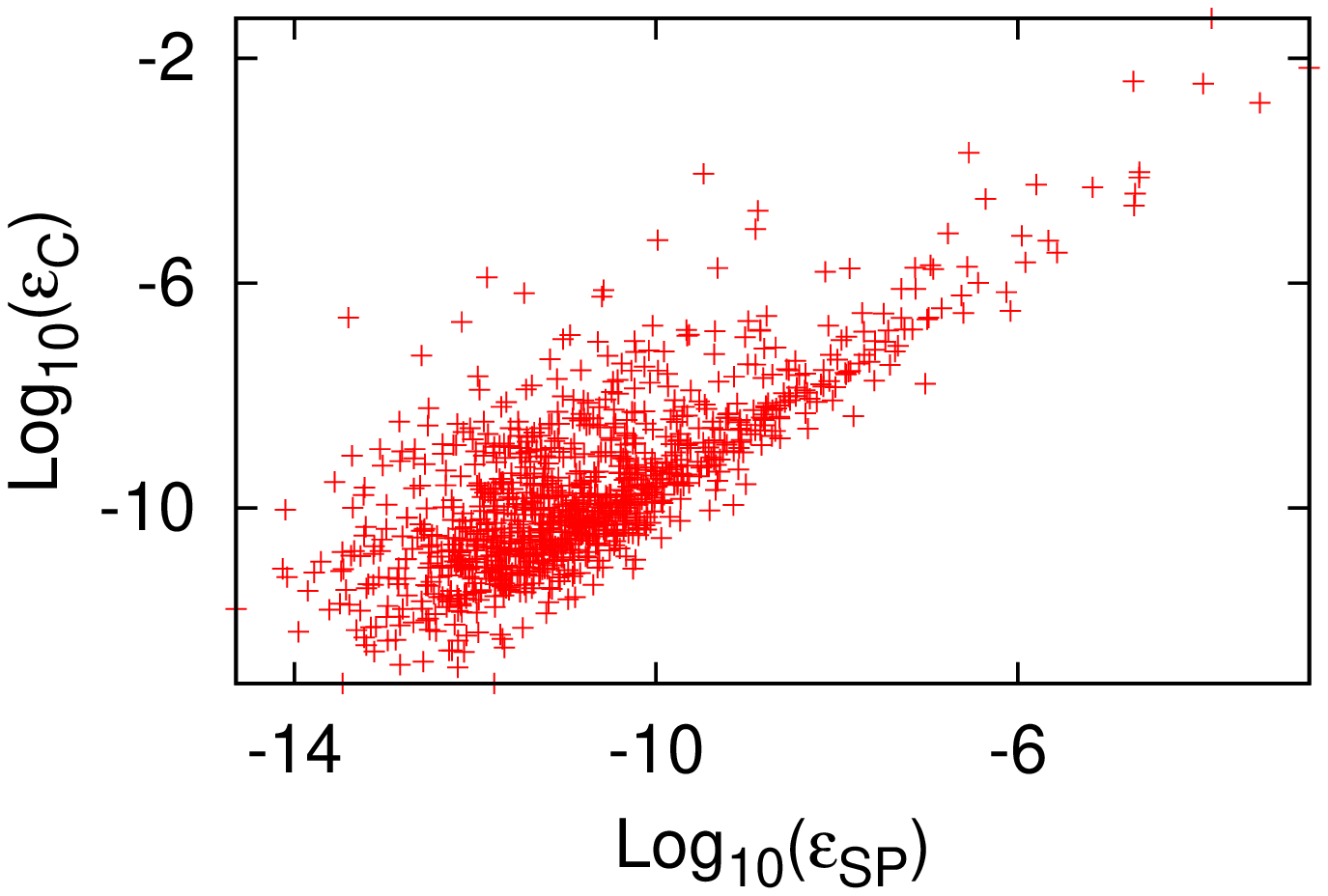}
\includegraphics[angle=270,scale=0.49]{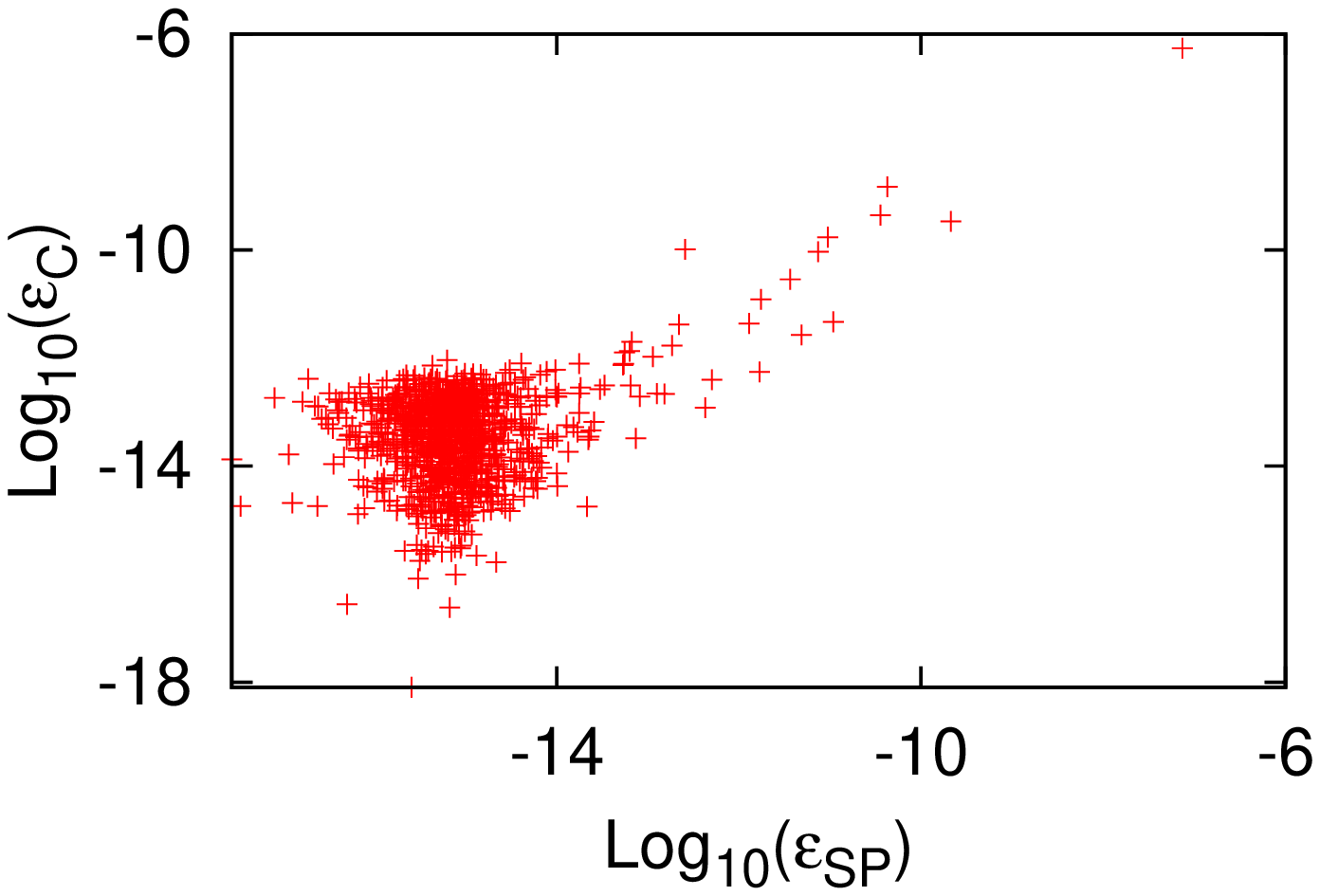}
\end{center}
\caption{Relative accuracy of the constant part
versus the accuracy on the single poles in double precision (left) and
quadrupole precision (right) for $N=6$ MHV amplitudes when sampling
1,000 phase space points.}
\label{fig:MHVN6}
\end{figure}

For the results shown in the above plots we choose to rerun all events
in quadrupole precision to get an overall picture. However, in
practice only a small fraction of phase space points require a
quadrupole precision treatment (this fraction can be read off the
plots in fig.~(\ref{fig:accuracy611}) and depends on $N$ and on the
target accuracy).
Therefore one needs a systematic procedure to decide which events
should be re-evaluated in quadrupole precision. 
One possible way is to verify the accuracy of the single poles
results.  The analytic single pole result is given in
eq.~(\ref{eq:poles}) for arbitrary number of gluons.  Since two-point
functions contain single poles, this checks the coefficients of the
two-point master integrals as well as the coefficients of the higher
point master integrals.
In fig.~(\ref{fig:MHVN6}) we investigate the correlation between the
accuracy of the single pole contribution and the constant part.
We plot the relative accuracy of the constant part $\log_{10}
(\epsilon_{\rm C})$ versus the accuracy on the single poles $\log_{10}
(\epsilon_{\rm SP})$ in double precision (left) for $N=6$ MHV
amplitudes when sampling 1,000 phase space points. The high
correlation between the accuracy of the constant part and the single
pole is evident.  In fig.~(\ref{fig:MHVN6}) (right) we show the
improvement when running the same points in quadrupole precision (note
the different scale on the y-axis).  This leaves us with a
straightforward estimate of the accuracy of the one-loop
evaluation. By comparing the numerical evaluated single pole result
against the analytic single pole result we can decide to switch to
quadruple precision to re-evaluate the full one-loop amplitude and get
the required precision.
Alternatively, one might choose to run in quadrupole precision
whenever any small denominator (for instance in the construction of
dual vectors) appears.
We did not fully investigate yet which method is the most efficient in
detecting potential instabilities. We will leave this to a future
study, but we anticipate that identifying dangerous phase space points
in order to increase the accuracy is not an issue.

\subsection{Time Dependence of the Algorithm}
\begin{figure}[t]
\begin{center}
\includegraphics[angle=270,scale=0.5]{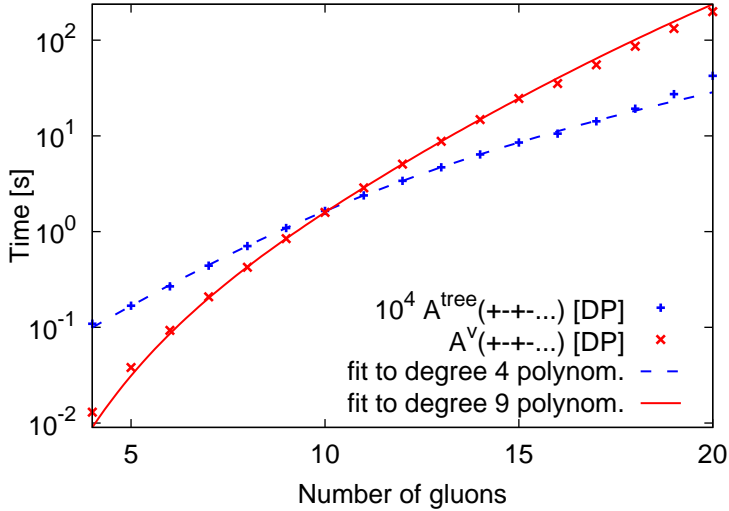}
\end{center}
\caption{Time in seconds needed to compute tree (blue, dashed) and
one-loop (red, solid) ordered amplitudes with gluons of alternating
helicity signs, $A_N^{[1]}(+-+-+...)$, as a function of the number of
external gluons ranging between 4 to 20 using a single 2.33 GHz Xeon
processor.  }
\label{fig:time}
\end{figure}
Compared to traditional Feynman diagram approaches, the power of this
method is that the time needed to compute one-loop amplitudes does not
grow factorially with the number of external legs. It is indeed quite
straightforward to estimate the scaling of time with the number of
gluons $N$. Within the so-called constructive implementation of
Berends-Giele recursion relations (or alternatively a recursive
implementation ``with memory'') the time required to compute tree
level ordered amplitudes grows as $\tau_{\rm tree, N}\propto
N^4$~\cite{Kleiss:1988ne}.
Altogether the number of tree amplitudes that one needs to
evaluate at one-loop is simply given by
\begin{eqnarray}
n_{\rm tree} &=& \left \{(D_{s_1}-2)^2 +(D_{s_2}-2)^2\right\}\\
&&\times\left(
5\, c_{5, \rm max}  {N \choose 5} +
4\, c_{4, \rm max}  {N \choose 4} +
3\, c_{3, \rm max}  {N \choose 3} +
2\, c_{2, \rm max}  \left[{N \choose 2} -N\right]\right)\nonumber\,,
\end{eqnarray}
where the first factor is due to the sum over polarization of the
internal cut gluons in $D_{s_1}$ and $D_{s_2}$ dimensions respectively
needed to determine the dimensional dependence of the one-loop
amplitude~\cite{GKM}.  The constants $c_{m, {\rm max}}$ denote the
number of times one needs to perform a multiple cut in order to fully
constraint the system of equations determining the master integral
coefficients. Explicitly we have $c_{5, \rm max} = 1$, $c_{4,
\rm max} = 5$, $c_{3, \rm max} = 10$, and $c_{2, \rm max} = 10$~\cite{GKM}. The
integer number in front counts the number of tree amplitudes per
multiple cut, finally the binomial coefficients corresponds to the
number of possible cuts (for two point functions we subtract the
vanishing contributions of the external self energy graphs).  From
this it follows that the time needed to evaluate a one-loop ordered
amplitude will for large $N$ scale as
\begin{equation}
\tau_{\rm one-loop, N} \sim n_{\rm tree} \cdot \tau_{\rm tree, N} \propto
N^9\,. 
\label{eq:timeoneloop}
\end{equation}
In fig.~(\ref{fig:time}) we plot the time needed to compute tree
(blue, dashed) and one-loop (red, solid) ordered amplitudes with
alternating helicity signs for the gluons, $A_N^{[1]}(+-+-\dots)$, as
a function of the number of gluons in the range between four and
twenty.  Time estimates refer to using a 2.33 GHz Xeon processor.
One can see that the times needed to compute tree and one-loop ordered
amplitudes are consistent with a $N^4$ and $N^9$ growth respectively
(we show a polynomial fit to those points as well).\footnote{The
evaluation times given in ref.~\cite{Berger:2008sj} for 6, 7 and 8
gluon MHV amplitudes make use of analytic expressions for the required
tree amplitudes. Nevertheless, the average time for the most
complicated 6 gluon helicity amplitudes quoted is 72ms per phase space
point, while we have an average computation time of 90ms for any of
the helicity amplitude.}
When running in quadrupole precision rather than in double precision
the evaluation time grows with a factor of approximately thirty.  We
verified that the scaling with $N$ is unchanged.

Finally we remark that the time needed to compute ``easier'' or more
``difficult'' helicity amplitudes is approximately the same, i. e.  we
checked that the plot looks essentially identical for other helicity
configurations.

\subsection{Results for fixed phase space points}
In this section we present sample results for one-loop helicity
amplitudes at fixed phase space points for $N =6,7$. Other results for
$N$ ranging from eight to twenty are given in Appendix~\ref{NumRes}.
Since the algorithm employed is independent of the chosen external
helicity vectors we show, apart from a comparison with known
amplitudes, the results for the ``most difficult'', alternating sign
helicity configurations, $A_N^{[1]}(+-+-\dots)$.  (A zero in the tables
means results smaller than $10^{-15}$.)

\paragraph{$N=6$}
We choose in this case the same phase space point as
in~\cite{Ellis:2006ss}, i.e. the six momenta $p_i$ are chosen as
follows, $p = (E,p_x,p_y,p_z),$\footnote{Note however that we use a
different convention to label the momentum components here.}
\begin{scriptsize}
\begin{eqnarray}
\label{specificpointN6}
p_ 1&=&(-3.000000000000000, 1.837117307087384,-2.121320343559642, 1.060660171779821)\nn
p_ 2&=&(-3.000000000000000,-1.837117307087384, 2.121320343559642,-1.060660171779821)\nn
p_ 3&=&( 2.000000000000000, 0.000000000000000,-2.000000000000000, 0.000000000000000)\nn
p_ 4&=&( 0.857142857142857, 0.000000000000000, 0.315789473684211, 0.796850604480708)\nn
p_ 5&=&( 1.000000000000000, 0.866025403784439, 0.184210526315789, 0.464829519280413)\nn
p_ 6&=&( 2.142857142857143,-0.866025403784439, 1.500000000000000,-1.261680123761121)\,.
\end{eqnarray}
\end{scriptsize}
We obtain the same results as in~\cite{Ellis:2006ss}, which we give in Table 1 for
completeness. 
\begin{scriptsize}
\TABLE[h]{
\begin{tabular}{|l|c|c|c|}
\hline
Helicity amplitude & $\cg/\epsilon^2$ & $\cg/\epsilon$ & 1 \\
\hline 
\hline 
$|A_6^{\rm tree}(++++++)|$ & - & - &  $1.767767365814634 \cdot 10^ {-15}$\\ 
$|A_{6}^{\rm v, unit}(++++++)|$ & $0$ & $0$ &  $ 0.529806483643855$ \\
$|A_{6}^{\rm v, num}(++++++)|$ & $1.060660419488780 \cdot 10^{-14}$&$  3.813284749527035 \cdot 10^{-14}$ & $0.529806483661295$\\
\hline
\hline
$|A_6^{\rm tree}(-+++++)|$ & - & - & $3.963158957208070\cdot 10^{-14}$ \\ 
$|A_{6}^{\rm v, unit}(-+++++)|$ & $1.011255761241711\cdot 10^{-11}$ &$6.753625348984687\cdot 10^{-10}$& $3.25996704351899$ \\
$|A_{6}^{\rm v, num}(-+++++)|$ &  $2.377895374324842 \cdot 10^{-13}$&$8.549005883762705 \cdot 10^{-13} $&  $3.25996705427236$ \\ 
\hline
\hline
$|A_6^{\rm tree}(--++++)|$ & - & - &   $28.4912816504432$\\ 
$|A_{6}^{\rm v, unit}(--++++)|$ & $   170.947689902659      $&$   614.590878376396  $&$   1373.74753500854      $\\
$|A_{6}^{\rm v, num}(--++++)|$ & $170.947689902659 $&$614.590878376397$ &$1373.74753500828$ \\ 
\hline
\hline
$|A_6^{\rm tree}(-+-+-+)|$ & - & - &  $3.13871539500808$\\ 
$|A_{6}^{\rm v, unit}(-+-+-+)|$ & $18.8322923700467 $&$67.7058293474830 $ &  $ 151.043950328960$ \\
$|A_{6}^{\rm v, num}(-+-+-+)|$ & $18.8322923700485$ & $67.7058292869577 $ & $151.043950337947$\\
\hline
\hline
$|A_6^{\rm tree}(+-+-+-)|$ & - & - &  $3.13871539500808 $\\ 
$|A_{6}^{\rm v, unit}(+-+-+-)|$ & $18.8322923700554 $&$67.7058292857048$ & $ 153.780101529836$ \\
$|A_{6}^{\rm v, num}(+-+-+-)|$ & $ 18.8322923700485 $ & $67.7058292869577  $ & $ 153.780101415986 $\\ 
\hline
\end{tabular}
\caption{Results for tree level and one-loop virtual (unrenormalized)
amplitudes in the FDH scheme for some helicity configurations for the
case of six external gluons for the phase space point of
eq.~(\ref{specificpointN6}). Comparison with known results is also
shown.}
}
\end{scriptsize}

\paragraph{$N=7$}
We randomly choose the following phase space point:
\begin{scriptsize}
\begin{eqnarray}
\label{specificpointN7}
p_ 1&=&(-3.500000000000000, 3.500000000000000, 0.000000000000000, 0.000000000000000)\nn
p_ 2&=&(-3.500000000000000,-3.500000000000000, 0.000000000000000, 0.000000000000000)\nn
p_ 3&=&( 1.721020317835363, 0.501455810979379, 0.991016354865482,-1.314663298501285)\nn
p_ 4&=&( 2.156731348769508,-1.341260229883955, 1.169853444323893, 1.218176516478751)\nn
p_ 5&=&( 1.037618172453970,-0.152260900991984,-0.998573413613423, 0.237316723937150)\nn
p_ 6&=&( 0.584066733036484, 0.353785124568635, 0.163503402186405,-0.435013415594666)\nn
p_ 7&=&( 1.500563427904675, 0.638280195327925,-1.325799787762357, 0.294183473680051)\nn
\end{eqnarray}
\end{scriptsize}
The results are given in Table 2.

\begin{scriptsize}
\TABLE[h]{
\begin{tabular}{|l|c|c|c|}
\hline
Helicity amplitude & $\cg/\epsilon^2$ & $\cg/\epsilon$ & 1 \\
\hline 
\hline 
$|A_7^{\rm tree}(+++++++)|$ & - & - &   $0$\\ 
$|A_{7}^{\rm v, unit}(+++++++)|$ & $  0 $&$  1.256534542409480\cdot 10^{-10} $&$  0.310169532972026      $\\
$|A_{7}^{\rm v, anly}(+++++++)|$ & $  0 $&$  1.250170111559883\cdot 10^{-15} $&$  0.310169533483183      $\\
\hline
\hline
$|A_7^{\rm tree}(-++++++)|$ & - & - & $0$ \\ 
$|A_{7}^{\rm v, unit}(-++++++)|$ & $  3.678212874319657\cdot 10^{-13} $&$  7.209572152581734\cdot 10^{-13} $&$  0.192052814810810      $\\
$|A_{7}^{\rm v, anly}(-++++++)|$ & $  2.713533399763100\cdot 10^{-15} $&$  8.924875144594874\cdot 10^{-15} $&$  0.192052814765395      $\\
\hline
\hline
$|A_7^{\rm tree}(--+++++)|$ & - & - & $2.10661283459449$ \\ 
$|A_{7}^{\rm v, unit}(--+++++)|$ &  $   14.7462898421614      $&$   48.5008939631214      $&$   87.3152155138790      $\\
$|A_{7}^{\rm v, anly}(--+++++)|$ &  $   14.7462898421614      $&$   48.5008939631213      $ &  $87.3152155138651$ \\
\hline
\hline
$|A_7^{\rm tree}(-+-+-+-)|$ & - & - &  $0.110186568094442$\\ 
$|A_{7}^{\rm v, unit}(-+-+-+-)|$ &  $  0.771305976661093      $&$   2.53684348996073      $&$   5.93361050294547     $\\
$|A_{7}^{\rm v, anly}(-+-+-+-)|$ &  $  0.771305976661095      $&$   2.53684348996075      $ & N.A. \\
\hline
\hline
$|A_7^{\rm tree}(+-+-+-+)|$ & - & - & $0.110186568094442$ \\ 
$|A_{7}^{\rm v, unit}(+-+-+-+)|$ & $  0.771305976661093      $&$   2.53684348996074      $&$ 6.04201240991614   $\\
$|A_{7}^{\rm v, anly}(+-+-+-+)|$ & $  0.771305976661095      $&$   2.53684348996075      $ & N.A. \\
\hline
\end{tabular}
\caption{Results for tree level and one-loop virtual (unrenormalized)
amplitudes in the FDH scheme for some helicity configurations for the
case of seven external gluons for the phase space point of
eq.~(\ref{specificpointN7}). Comparison with analytical results, when
available, is also shown.}  }
\end{scriptsize}


\section{Conclusions and outlook}
\label{sec:conclu}
In this paper we present the numerical implementation of an integer
dimensional on-shell method for calculating one-loop amplitudes. The
method used was developed in refs.~\cite{EGK,GKM} which we followed
closely.  The resulting program, {\bf Rocket}, is a {\tt Fortran} 95
code.  The only limitation on the number of external particles are the
available computer resources.
   
As a first example and test of the implemented algorithm we calculated
purely gluonic color ordered one-loop amplitudes. We explicitly study
the properties and time-behavior of the algorithm up to twenty
external gluons.  The scaling of the computer time needed to evaluate
color ordered one-loop amplitudes is consistent with the theoretical
estimate of a rank nine polynomial.  Comparisons to existing analytic
results for special helicity configuration shows a good numerical
accuracy and only for a limited set of events quadrupole precision is
required.  For completeness and later reference we give numerical
results for explicit events.

Now that the algorithm has been validated up to twenty external
particles we plan to include internal and external (massive) quarks
and external vector bosons.  This allows us to compute one-loop
amplitudes to a large range of processes relevant for both the LHC and
Tevatron experiments.

We envision the collection of one-loop matrix elements to be
integrated through matching into parton shower Monte Carlo's. The
parton shower Monte Carlo will integrate the one-loop matrix elements
with the radiative contributions.  This will allow a seamless integration
of the available tools in a single framework and will provide a
complete and advanced analysis tool for experimentalists.

\acknowledgments We thank Fabio Maltoni for providing tree level
amplitudes for checks and Lance Dixon and Keith Ellis for comments on
the manuscript. W.G thanks Jan Winter for useful discussions.
 G.Z. is supported by the British Science and
Technology Facilities Council (STFC). G.Z. would like to thank Mainz
University and CERN for hospitality while part of this work was carried
out. 

\newpage


\appendix

\section{More numerical results}
\label{NumRes}

In this appendix we list some explicit results for high multiplicity
events.

\paragraph{$N=8$}
We randomly choose the following phase space point:
\begin{scriptsize}
\begin{eqnarray}
\label{specificpointN8}
p_ 1&=&(-4.000000000000000, 4.000000000000000, 0.000000000000000, 0.000000000000000)\nn
p_ 2&=&(-4.000000000000000,-4.000000000000000, 0.000000000000000, 0.000000000000000)\nn
p_ 3&=&( 1.449692512284710, 0.958721567196264, 0.596442789329140, 0.909239977027169)\nn
p_ 4&=&( 1.009340416556955,-0.511164560265637, 0.004406444973267, 0.870321464785572)\nn
p_ 5&=&( 1.065731003301513, 0.640468049755497,-0.851783840257115,-0.006894789139810)\nn
p_ 6&=&( 1.402207989626767, 0.338467194877458,-0.023915684844839,-1.360534911049079)\nn
p_ 7&=&( 1.702524230799814,-0.823031265059939, 1.480876536483203,-0.167967189914708)\nn
p_ 8&=&( 1.370503847430242,-0.603460986503644,-1.206026245683657,-0.244164551709144)\,.
\end{eqnarray}
\end{scriptsize}
The results are given in Table 3.

\begin{scriptsize}
\TABLE[h]{
\begin{tabular}{|l|c|c|c|}
\hline
Helicity amplitude & $\cg/\epsilon^2$ & $\cg/\epsilon$ & 1 \\
\hline 
\hline 
$|A_8^{\rm tree}(++++++++)|$ & - & - & $ 0$ \\ 
$|A_{8}^{\rm v, unit}(++++++++)|$ & $0$ &$0$ & $ 0.196700600695691      $\\
$|A_{8}^{\rm v, anly}(++++++++)|$ & $  3.853462894343397\cdot 10^{-15} $&$  1.441159379540454\cdot 10^{-14} $&$  0.196700600738201      $\\
\hline
\hline
$|A_8^{\rm tree}(-+++++++)|$ & - & - & $ 2.257277386254959\cdot 10^{-15}$ \\ 
$|A_{8}^{\rm v, unit}(-+++++++)|$ &  $  0 $&$  1.965638104048654\cdot 10^{-10} $&$  0.528774716493063      $\\
$|A_{8}^{\rm v, anly}(-+++++++)|$ & $  1.805821909003967\cdot 10^{-14} $&$  6.753606439965886\cdot 10^{-14} $&$  0.528774717652170      $\\
\hline
\hline
$|A_8^{\rm tree}(--++++++)|$ & - & - &  $4.33318919466960$ \\ 
$|A_{8}^{\rm v, unit}(--++++++)|$ &   $   34.6655135573561      $&$   129.645805347145      $&$   274.299773434926      $\\
$|A_{8}^{\rm v, anly}(--++++++)|$ &   $   34.6655135573568      $&$   129.645805291409      $ &  $274.299773434900$\\
\hline
\hline
$|A_8^{\rm tree}(-+-+-+-+)|$ & - & - & $7.261522613885579\cdot 10^{-2}$ \\ 
$|A_{8}^{\rm v, unit}(-+-+-+-+)|$ &  $  0.580921809110773      $&$   2.17259368023597      $&$    5.47630381976679    $\\
$|A_{8}^{\rm v, anly}(-+-+-+-+)|$ &   $  0.580921809110846      $&$   2.17259368244769      $ & N.A. \\
\hline
\hline
$|A_8^{\rm tree}(+-+-+-+-)|$ & - & - &  $7.261522613885579\cdot 10^{-2}$\\
$|A_{8}^{\rm v, unit}(+-+-+-+-)|$ & $  0.580921809110862      $&$   2.17259368781042      $&$   4.92550054630729      $\\
$|A_{8}^{\rm v, anly}(+-+-+-+-)|$ & $  0.580921809110846      $&$   2.17259368244769      $ & N.A. \\
\hline
\end{tabular}
\caption{Results for tree level and one-loop virtual (unrenormalized)
amplitudes in the FDH scheme for some helicity configurations for the
case of eight external gluons for the phase space point of
eq.~(\ref{specificpointN8}). Comparison with analytical results, when
available, is also shown.}  }
\end{scriptsize}

\newpage

\paragraph{$N=9$}
We randomly choose the following phase space point:
\begin{scriptsize}
\begin{eqnarray}
\label{specificpointN9}
p_ 1&=&(-4.500000000000000, 4.500000000000000, 0.000000000000000, 0.000000000000000)\nn
p_ 2&=&(-4.500000000000000,-4.500000000000000, 0.000000000000000, 0.000000000000000)\nn
p_ 3&=&( 0.837513535184208,-0.699991554911619,-0.341245468610472, 0.308208168000110)\nn
p_ 4&=&( 0.173971130750340,-0.127314067058472,-0.036582419646361, 0.112777698311325)\nn
p_ 5&=&( 1.527678295320022,-0.329915424080283,-1.066446319295822,-1.042904135098816)\nn
p_ 6&=&( 1.087863978393825,-0.427338530001958,-0.959598549771965, 0.282843489474564)\nn
p_ 7&=&( 2.837576052736588, 0.906765089674632, 2.135787640996821,-1.633409342380760)\nn
p_ 8&=&( 2.065394708000926, 0.474995872769939, 0.683997980573425, 1.890074332734359)\nn
p_ 9&=&( 0.470002299614090, 0.202798613607760,-0.415912864245626, 0.082409788959218).
\end{eqnarray}
\end{scriptsize}
The results are given in Table 4.

\begin{scriptsize}
\TABLE[h]{
\begin{tabular}{|l|c|c|c|}
\hline
Helicity amplitude & $\cg/\epsilon^2$ & $\cg/\epsilon$ & 1 \\
\hline 
\hline
$|A_9^{\rm tree}(+++++++++)|$ & - & - & $2.992915640032351\cdot 10^{-14}$ \\ 
$|A_{9}^{\rm v, unit}(+++++++++)|$ & $  4.860269836292316\cdot 10^{-12} $&$  1.845193695700690\cdot 10^{-08} $&$   5.66655561706295      $\\
$|A_{9}^{\rm v, anly}(+++++++++)|$ & $  2.693624076029116\cdot 10^{-13} $&$  1.176695244346755\cdot 10^{-12} $&$   5.66655558047311      $\\
\hline
\hline
$|A_9^{\rm tree}(-++++++++)|$ & - & - & $9.114087930248735 \cdot 10^{-14}$ \\ 
$|A_{9}^{\rm v, unit}(-++++++++)|$ &  $  3.938371378126140\cdot 10^{-11} $&$  2.340429860576292\cdot 10^{-08} $&$   1.06208646061428      $\\
$|A_{9}^{\rm v, anly}(-++++++++)|$ &  $  8.202679137223861\cdot 10^{-13} $&$  3.583296428617654\cdot 10^{-12} $&$   1.06208646798175      $\\
\hline
\hline
$|A_9^{\rm tree}(--+++++++)|$ & - & - & $32.3229667945508$ \\ 
$|A_{9}^{\rm v, unit}(--+++++++)|$ & $   290.906701096922      $&$   1270.81033486132      $&$  3625.43061670521      $\\
$|A_{9}^{\rm v, anly}(--+++++++)|$ & $   290.906701150957      $&$1270.81033630185      $ & $3625.43061670594$ \\
\hline
\hline
$|A_9^{\rm tree}(-+-+-+-+-)|$ & - & - & $0.453521966367950$ \\ 
$|A_{9}^{\rm v, unit}(-+-+-+-+-)|$   & $ 4.08169769666186      $&$   17.8306776720814      $&$  57.1063950462874  $\\
$|A_{9}^{\rm v, anly}(-+-+-+-+-)|$   & $ 4.08169769731155      $&$   17.8306776807844      $ & N.A. \\
\hline
\hline
$|A_9^{\rm tree}(+-+-+-+-+)|$ & - & - & $0.453521966367950$ \\ 
$|A_{9}^{\rm v, unit}(+-+-+-+-+)|$ & $ 4.08169769662055      $&$   17.8306776454842      $&$   55.0153807707576 $\\
$|A_{9}^{\rm v, anly}(+-+-+-+-+)|$ & $ 4.08169769731155      $&$   17.8306776807844      $ & N.A. \\
\hline
\end{tabular}
\caption{Results for tree level and one-loop virtual (unrenormalized)
amplitudes in the FDH scheme for some helicity configurations for the
case of nine external gluons for the phase space point of
eq.~(\ref{specificpointN9}). Comparison with analytical results, when
available, is also shown.}}
\end{scriptsize}

\newpage
\paragraph{$N=10$}
We randomly choose the following phase space point:
\begin{scriptsize}
\begin{eqnarray}
\label{specificpointN10}
p_ 1&=&(-5.000000000000000, 5.000000000000000, 0.000000000000000, 0.000000000000000)\nn
p_ 2&=&(-5.000000000000000,-5.000000000000000, 0.000000000000000, 0.000000000000000)\nn
p_ 3&=&( 1.532520310665362, 1.513070043279136,-0.187195319404816, 0.155548896254710)\nn
p_ 4&=&( 0.258067002946760,-0.183609750404875,-0.180504757414873, 0.017437606394811)\nn
p_ 5&=&( 0.933615908667822,-0.324162340665056,-0.045792611756259,-0.874334305926935)\nn
p_ 6&=&( 1.406380992739491, 0.945471671828880,-0.641345241905772,-0.820162846752305)\nn
p_ 7&=&( 0.826140594064319, 0.546878011685178,-0.615447980566360,-0.068238586686725)\nn
p_ 8&=&( 1.003938368361987,-0.166638867622269, 0.778081660907425,-0.612137782060901)\nn
p_ 9&=&( 0.626103283169757,-0.367431900674052, 0.140872898707371,-0.486984543874634)\nn
p_{10}&=&( 3.413233539384504,-1.963576867426942, 0.751331351433283, 2.688871562651979).
\end{eqnarray}
\end{scriptsize}
The results are given in Table 5.
\begin{scriptsize}
\TABLE[h]{
\begin{tabular}{|l|c|c|c|}
\hline
Helicity amplitude & $\cg/\epsilon^2$ & $\cg/\epsilon$ & 1 \\
\hline 
\hline
$|A_{10}^{\rm tree}(++++++++++)|$ & - & - & $7.645214091184737\cdot 10^{-14}$\\ 
$|A_{10}^{\rm v, unit}(++++++++++)|$ &  $  2.616999209810146\cdot 10^{-13} $&$  7.453142378465002\cdot 10^{-07} $&$   18.4349011284670      $\\
$|A_{10}^{\rm v, anly}(++++++++++)|$ &  $  7.645214091184737\cdot 10^{-13} $&$  3.853184186191476\cdot 10^{-12} $&$   18.4349011284671      $\\
\hline
\hline
$|A_{10}^{\rm tree}(-+++++++++)|$ & - & - & $3.138928592085274\cdot 10^{-13}$\\ 
$|A_{10}^{\rm v, unit}(-+++++++++)|$ &   $  1.729567134060808\cdot 10^{-11} $&$  3.462486730362966\cdot 10^{-06} $&$ 14.1180690283674     $\\
$|A_{10}^{\rm v, anly}(-+++++++++)|$ &   $  3.138928592085274\cdot 10^{-12} $&$  1.582018484813023\cdot 10^{-11} $&$   14.1180690283692      $\\
\hline
\hline
$|A_{10}^{\rm tree}(--++++++++)|$ & - & - & $489.972695666341$\\
$|A_{10}^{\rm v, unit}(--++++++++)|$ &  $   4899.72695665607      $&$   24694.6000400099      $&$   75844.9101458089    $\\
$|A_{10}^{\rm v, anly}(--++++++++)|$ &  $   4899.72695666341      $&$24694.6000476827      $ &  $75844.9101457814$ \\
\hline
\hline
$|A_{10}^{\rm tree}(-+-+-+-+-+)|$ & - & - & $9.34611372008902$\\
$|A_{10}^{\rm v, unit}(-+-+-+-+-+)|$   &  $   93.4611371998759      $&$   471.043678702711      $&$  1481.27447605664 $\\
$|A_{10}^{\rm v, anly}(-+-+-+-+-+)|$   &  $   93.4611372008902      $&$   471.043677247939      $ & N.A. \\
\hline
\hline
$|A_{10}^{\rm tree}(+-+-+-+-+-)|$ & - & - & $9.34611372008902$\\
$|A_{10}^{\rm v, unit}(+-+-+-+-+-)|$ &  $   93.4611371995618      $&$   471.043674005742      $&$  1503.97025803111  $\\
$|A_{10}^{\rm v, anly}(+-+-+-+-+-)|$ &  $   93.4611372008902      $&$   471.043677247939      $ & N.A. \\
\hline
\end{tabular}
\caption{Results for tree level and one-loop virtual (unrenormalized)
amplitudes in the FDH scheme for some helicity configurations for the
case of ten external gluons for the phase space point of
eq.~(\ref{specificpointN10}). Comparison with analytical results, when
available, is also shown.}
}
\end{scriptsize}

\newpage

\paragraph{$N=15$}
We randomly choose the following phase space point:
\begin{scriptsize}
\begin{eqnarray}
p_1&=&(-7.500000000000000, 7.500000000000000, 0.000000000000000, 0.000000000000000)\nn
p_2&=&(-7.500000000000000,-7.500000000000000, 0.000000000000000, 0.000000000000000)\nn
p_3&=&( 0.368648489648050, 0.161818085189973, 0.125609635286264,-0.306494430207942)\nn
p_4&=&( 0.985841964092509,-0.052394238926518,-0.664093578996812, 0.726717923425790)\nn
p_5&=&( 1.470453194926588,-0.203016239158633, 0.901766792550452,-1.143605551298596)\nn
p_6&=&( 2.467058579094687,-1.840106401193462, 0.715811527707121, 1.479189075734789)\nn
p_7&=&( 0.566021478235079,-0.406406330753485,-0.393435666409983,-0.020556861225509)\nn
p_8&=&( 0.419832726637289,-0.214182754609525, 0.074852807863799,-0.353245414886707)\nn
p_9&=&( 2.691168687878469, 1.868400546247601, 1.850615607221259,-0.571568175905795)\nn
p_{10}&=&( 1.028090983779864,-0.986442664896249,-0.193408556327968, 0.215627155388572)\nn
p_{11}&=&( 1.377779821947130,-0.155359745837053,-1.074009172530291,-0.848908054184264)\nn
p_{12}&=&( 1.432526153404585, 0.621168997409793,-0.290964068761809, 1.257624811911176)\nn
p_{13}&=&( 0.335532948820133, 0.244811479043329, 0.138986808214636, 0.182571538348285)\nn
p_{14}&=&( 1.085581415795683, 0.330868645896313,-0.756382142822373,-0.704910635118478)\nn
p_{15}&=&( 0.771463555739934, 0.630840621587917,-0.435349992994295, 0.087558618018677).
\label{specificpointN15}
\end{eqnarray}
\end{scriptsize}
The results are given in Table 6.

\begin{scriptsize}
\TABLE[h]{
\begin{tabular}[[h]{|l|c|c|c|}
\hline
Helicity amplitude & $\cg/\epsilon^2$ & $\cg/\epsilon$ & 1 \\
\hline 
\hline
$|A_{15}^{\rm tree}(++++\dots)|$ & - & - & $0$\\ 
$|A_{15}^{\rm v, unit}(++++\dots)|$ & $  0 $&$  0 $&$ 1.07572071884782      $\\
$|A_{15}^{\rm v, anly}(++++\dots)|$ & $  0 $&$  0 $&$ 1.07572071880769      $\\
\hline
\hline
$|A_{15}^{\rm tree}(-+++\dots++)|$ & - & - & $ 0$\\ 
$|A_{15}^{\rm v, unit}(-+++\dots++)|$ &  $  0 $&$  0 $&$  0.181194659968483      $\\
 $|A_{15}^{\rm v, anly}(-+++\dots++)|$ & $  0 $&$  0 $&$  0.181194659846677      $\\
 \hline
\hline
$|A_{15}^{\rm tree}(--+++\dots++)|$ & - & - & $7.45782101450887$\\
$|A_{15}^{\rm v, unit}(--++\dots++)|$ &  $   111.867315217633      $&$   586.858955605213      $&$   1810.13038312828      $\\
$|A_{15}^{\rm v, anly}(--++\dots++)|$ &  $   111.867315217633      $&$   586.858955605213      $&$   1810.13038312852      $\\
\hline
\hline
$|A_{15}^{\rm tree}(-+-\dots+-)|$ & - & - & $5.851039428822597 \cdot 10^{-3}$\\
$|A_{15}^{\rm v, unit}(-+-\dots+-)|$   &  $  8.776559143021942 \cdot 10^{-2} $&$  0.460420629357800      $&$   1.52033417713680      $\\
$|A_{15}^{\rm v, anly}(-+-\dots+-)|$   &  $  8.776559143233895 \cdot 10^{-2} $&$  0.460420661976678      $ & N.A. \\
\hline
\hline
$|A_{15}^{\rm tree}(+-+\dots-+)|$ & - & - & $5.851039428822597 \cdot 10^{-3}$\\
$|A_{15}^{\rm v, unit}(+-+\dots-+)|$ & $  8.776559143021942 \cdot 10^{-2} $&$  0.460420565320471      $&$   1.52960647292231      $\\
$|A_{15}^{\rm v, anly}(+-+\dots-+)|$ &  $  8.776559143233895 \cdot 10^{-2} $&$  0.460420661976678      $ & N.A. \\
\hline
\end{tabular}
\caption{Results for tree level and one-loop virtual (unrenormalized)
amplitudes in the FDH scheme for some helicity configurations for the
case of fifteen external gluons for the phase space point of
eq.~(\ref{specificpointN15}). Comparison with analytical results, when
available, is also shown. The present results have been obtained by
running in quadrupole precision.}}
\end{scriptsize}

\newpage 

\paragraph{$N=20$}
We randomly choose the following phase space point:
\begin{scriptsize}
\begin{eqnarray}
p_ 1&=&(-10.00000000000000,10.000000000000000, 0.000000000000000, 0.000000000000000)\nn
p_ 2&=&(-10.00000000000000,-10.00000000000000, 0.000000000000000, 0.000000000000000)\nn
p_ 3&=&( 0.325540699096246,-0.312416065575525, 0.033104753113790, 0.085305474968826)\nn
p_ 4&=&( 0.848759269032669,-0.024590847182333,-0.848289178399890, 0.013894488596740)\nn
p_ 5&=&( 1.570982650317940, 0.440773273547400, 0.720246141672893,-1.324745599853646)\nn
p_ 6&=&( 0.553167263468587,-0.230303833105897,-0.480156399163221,-0.149679651832391)\nn
p_ 7&=&( 0.503893441998193,-0.080227334603420,-0.333549326254518, 0.369075903611134)\nn
p_ 8&=&( 1.342531690799994,-0.248744151369669, 0.500198592589739, 1.220786244980225)\nn
p_ 9&=&( 2.116396457930369, 0.026327408340262,-0.849622753763036,-1.938190395961762)\nn
p_{10}&=&( 0.602748352923314,-0.444717464828695,-0.325013925746484, 0.244740477851469)\nn
p_{11}&=&( 1.497270156443179,-1.005250846935450, 0.200161613080160,-1.091432079774134)\nn
p_{12}&=&( 1.403440100824101, 0.989809299131399, 0.670534066899850,-0.735054918411486)\nn
p_{13}&=&( 1.968885859795150, 1.683278410651605,-0.987433952876609,-0.260882176167652)\nn
p_{14}&=&( 0.537434314204394, 0.448194620535905,-0.219499970243363, 0.199441688897247)\nn
p_{15}&=&( 2.339779276321334,-1.532082130972216, 0.290600646542316, 1.744374578491559)\nn
p_{16}&=&( 0.894504093025053,-0.210636900721777, 0.805755963464840, 0.326384735908015)\nn
p_{17}&=&( 0.306394773317926,-0.289150435585858, 0.066148943112216,-0.076773042418471)\nn
p_{18}&=&( 0.560235842911576,-0.343584006920049, 0.286594014336811, 0.337161831792864)\nn
p_{19}&=&( 1.070093313805907, 0.323329544655228, 1.003388464788536,-0.183764236276963)\nn
p_{20}&=&( 1.557942443784069, 0.809991460939089,-0.533167693154030, 1.219356675598426)\nn
\label{specificpointN20}
\end{eqnarray}
\end{scriptsize}
The results are given in Table 7.

\begin{scriptsize}
\TABLE[h]{
\begin{tabular}{|l|c|c|c|}
\hline
Helicity amplitude & $\cg/\epsilon^2$ & $\cg/\epsilon$ & 1 \\
\hline 
\hline
$|A_{20}^{\rm tree}(++++\dots)|$ & - & - & $  0$\\ 
$|A_{20}^{\rm v, unit}(++++\dots)|$      & $  0 $&$  0 $&$   1.78947750851720      $\\
$|A_{20}^{\rm v, anly}(++++\dots)|$      & $  0 $&$  0 $&$   1.789477509283  $\\
\hline
\hline
$|A_{20}^{\rm tree}(-+++\dots++)|$ & - & - & $  0$\\ 
$|A_{20}^{\rm v, unit}(-+++\dots++)|$      & $  0 $&$  0 $&$  0.303337144522210      $\\
$|A_{20}^{\rm v, anly}(-+++\dots++)|$      & $  0 $&$  0 $&$  0.303337141901917 $\\
\hline
\hline
$|A_{20}^{\rm tree}(--+++\dots++)|$ & - & - & $ 16.7096151501841$\\
$|A_{20}^{\rm v, unit}(--++\dots++)|$ &  $   334.192303003683      $&$   1995.15970325579      $&$   6882.49682704505      $\\
$|A_{20}^{\rm v, anly}(--++\dots++)|$ & $   334.192303003683      $&$   1995.15970325579      $&$   6882.49682704481      $\\
\hline
\hline
$|A_{20}^{\rm tree}(-+-\dots+-)|$ & - & - & $2.0970621000196\cdot 10^{-5}$\\
$|A_{20}^{\rm v, unit}(-+-\dots+-)|$   & $  4.194124200605681\cdot 10^{-4} $&$  2.503931965487835\cdot 10^{-3} $&$  8.456871985787404\cdot 10^{-3} $\\
$|A_{20}^{\rm v, anly}(-+-\dots+-)|$   & $  4.194124200039273\cdot 10^{-4} $&$ 2.503931873702081\cdot 10^{-3} $ & N.A. \\
\hline
\hline
$|A_{20}^{\rm tree}(+-+\dots-+)|$ & - & - & $2.0970621000196\cdot 10^{-5}$\\
$|A_{20}^{\rm v, unit}(+-+\dots-+)|$ & $  4.194124200605681\cdot 10^{-4} $&$  2.503931902332899\cdot 10^{-3} $&$  9.203156962017870\cdot 10^{-3} $\\
$|A_{20}^{\rm v, anly}(+-+\dots-+)|$ & $  4.194124200039273\cdot 10^{-4} $&$ 2.503931873702081\cdot 10^{-3} $ & N.A. \\
\hline
\end{tabular}
\caption{Results for tree level and one-loop virtual (unrenormalized)
amplitudes in the FDH scheme for some helicity configurations for the
case of twenty external gluons for the phase space point of
eq.~(\ref{specificpointN20}). Comparison with analytical results, when
available, is also shown. The present results have been obtained by
running in quadrupole precision.}}
\end{scriptsize}


\end{document}